\documentclass[conference]{IEEEtran}
\IEEEoverridecommandlockouts

\usepackage{cite}
\usepackage{amsmath,amssymb,amsfonts}
\usepackage{algorithmic}
\usepackage{graphicx}
\usepackage{textcomp}
\usepackage{xcolor}
\usepackage{booktabs}
\usepackage{url}
\def\BibTeX{{\rm B\kern-.05em{\sc i\kern-.025em b}\kern-.08em
    T\kern-.1667em\lower.7ex\hbox{E}\kern-.125emX}}

\begin{document}

\title{BanglaFake: Constructing and Evaluating a Specialized Bengali Deepfake Audio Dataset}

\author{

\IEEEauthorblockN{Istiaq Ahmed Fahad}
\IEEEauthorblockA{
Institute of Information Technology\\
University of Dhaka\\
Dhaka, Bangladesh\\
bsse1204@iit.du.ac.bd
}
\and

\IEEEauthorblockN{Kamruzzaman Asif}
\IEEEauthorblockA{
Institute of Information Technology\\
University of Dhaka\\
Dhaka, Bangladesh\\
bsse1217@iit.du.ac.bd
}
\and

\IEEEauthorblockN{Sifat Sikder}
\IEEEauthorblockA{
Institute of Information Technology\\
University of Dhaka\\
Dhaka, Bangladesh\\
bsse1221@iit.du.ac.bd
}

}

\maketitle

\begin{abstract}
Deepfake audio detection is challenging for low-resource languages like Bengali due to limited datasets and subtle acoustic features. To address this, we introduce \textbf{BanglaFake}, a Bengali Deepfake Audio Dataset with \textbf{13,796 real} and \textbf{12,796 deepfake} utterances. Synthetic speech is generated using \textbf{SOTA} Text-to-Speech (\textbf{TTS}) models, ensuring high naturalness and quality. We evaluate the dataset through both \textbf{qualitative} and \textbf{quantitative} analyses. \textit{Mean Opinion Score (MOS)} from 30 native speakers shows \textbf{Robust-MOS} of \textbf{3.40} (naturalness) and \textbf{4.01} (intelligibility). \textit{t-SNE visualization} of \textbf{MFCCs} highlights real vs. fake differentiation challenges. This dataset serves as a crucial resource for advancing deepfake detection in Bengali, addressing the limitations of low-resource language research.
\end{abstract}

\begin{IEEEkeywords}
speech synthesis, deepfake, Bangla TTS, Bangla deepfake corpus
\end{IEEEkeywords}

\section{Introduction}
Deepfake audio generation presents a growing challenge for speech processing and biometric verification systems. Automatic speaker verification (ASV) tools, widely used for biometric identification, are increasingly vulnerable to spoofing attacks, particularly deepfake audios. Since the term `deepfake' was coined in 2017, AI-driven manipulation of audio and video has enabled the creation of hyper-realistic fake content, posing significant security risks \cite{bitesize2019deepfakes}. Deepfake audios exploit advanced speech synthesis and voice conversion techniques to mimic a speaker's voice with high fidelity. These capabilities have been misused for fraudulent activities, underscoring the necessity for robust detection mechanisms. However, effective countermeasures depend on the availability of large, high-quality datasets \cite{azeemi2022dataset}, which remain scarce for low-resource languages like Bengali.

In this paper, we introduce BanglaFake, the first publicly available Bengali deepfake audio dataset. Our dataset serves as a benchmark for detecting deepfake audios in Bengali, aiming to enhance deepfake detection research for low-resource languages and strengthen safeguards against AI-generated media misuse.

\subsection{Contributions}
Our research makes the following key contributions:

\begin{itemize}
\item We present \textbf{BanglaFake}, a novel Bengali deepfake audio dataset containing \textbf{13,796} real utterances and \textbf{12,796} deepfake utterances, specifically designed for training deepfake detection models. The dataset is publicly available on Hugging Face\footnote{\url{https://huggingface.co/datasets/sifat1221/banglaFake}}. Furthermore, all relevant artifacts are available in our GitHub repository\footnote{\url{https://github.com/KamruzzamanAsif/BanglaFake}}.

\item We generate high-quality Bengali deepfake audio samples using a state-of-the-art end-to-end TTS model based on the \textbf{VITS} technique, trained from scratch for this task.

\item We evaluate the dataset using human assessments (Mean Opinion Score, MOS) across two key dimensions: \textbf{naturalness} and \textbf{clarity}. Additionally, we employ \textbf{t-SNE} visualization to illustrate the separability between real and deepfake audio samples, leveraging its ability to reduce high-dimensional data into comprehensible visual representations.
\end{itemize}

\section{Related Work}

\subsection{Deepfake Detection Models}
Audio deepfake detection has advanced significantly, leveraging machine learning techniques to distinguish between real and synthetic speech. Traditional methods rely on handcrafted feature extraction, such as Mel-Frequency Cepstral Coefficients (\textbf{MFCCs}) and spectrogram-based analysis, combined with classifiers like Support Vector Machines (\textbf{SVM}) and Gaussian Mixture Models (\textbf{GMM}) \cite{tomashenko2020exploring}. However, recent advancements in deep learning have improved detection and enabled the generation of highly realistic synthetic speech. Techniques such as Variational Autoencoders (\textbf{VAEs}) \cite{kingma2019introduction}, Generative Adversarial Networks (GANs), and transformer-based models like \textbf{FastSpeech} \cite{ren2019fastspeech} have significantly enhanced the quality and naturalness of deepfake audio. Consequently, detection models now leverage convolutional neural networks (CNNs), recurrent neural networks (RNNs), and transformer-based architectures such as \textbf{wav2vec 2.0} \cite{baevski2020wav2vec} to effectively identify subtle artifacts and inconsistencies present in synthetic speech. Besides recent approaches incorporate \textbf{self-supervised learning} to improve detection robustness, particularly in low-resource languages \cite{gong2024zmm}. Additionally, zero-shot deepfake detection has emerged, allowing models trained on high-resource languages to generalize to low-resource settings \cite{saeki2023learning}.

\subsection{Deepfake Detection Datasets}
Benchmark datasets like \textbf{ASVspoof} \cite{yamagishi2021asvspoof} and the \textbf{ADD challenge} \cite{yi2023add} have driven progress in this field, but they primarily focus on high-resource languages like English and Chinese. To mitigate this language barrier recent studies propose cross-lingual data augmentation for deepfake detection in underrepresented languages \cite{liu2023comparative}. Moreover, the \textbf{ZMM-TTS framework} has demonstrated self-supervised speech representations for improving deepfake detection in unseen languages \cite{gong2024zmm}. Also the FakeAVCeleb \cite{khalid2021fakeavceleb} dataset expands detection research by including multi-modal deepfake content.

\subsection{Low-Resource Language TTS and Deepfake Generation}
Advancements in low-resource TTS leverage cross-lingual transfer learning \cite{saeki2023learning} and low-latency zero-shot TTS \cite{dang2024livespeech} to synthesize high-quality speech with minimal training data. Additionally, neural audio codecs and generative models have enabled zero-shot TTS, improving synthetic speech quality \cite{wang2023neural}. These advancements underscore the need for specialized datasets to keep detection models effective.

Our work addresses this gap by introducing BanglaFake, the \textbf{first dedicated Bengali deepfake audio dataset}. Unlike general Text-to-Speech (TTS) datasets, which are optimized for speech synthesis rather than detection, BanglaFake is specifically curated to enhance the robustness of deepfake detection models. Incorporating high-quality real and synthetic utterances, it provides a valuable benchmark for evaluating deepfake detection techniques for the Bengali language.

\section{Dataset Description}

The \textbf{BanglaFake} dataset is designed to support developing and evaluating deepfake audio detection models specifically for the Bengali language. It includes both real and synthetic Bengali speech samples, providing a comprehensive resource for training and testing models in the field of audio forensics and deepfake detection.

\subsection{Dataset Overview}
The \textbf{BanglaFake} dataset consists of real and synthetic (deepfake) Bengali speech samples. It is one of the first datasets to address the issue of deepfake audio detection in Bengali, and it serves as a foundation for evaluating detection models aimed at identifying manipulated audio.

\begin{table}[htbp]
\caption{Overview of the BanglaFake Dataset}
\label{tab:banglafake_dataset_overview}
\centering
\begin{tabular}{|l|c|c|}
\hline
\textbf{Category} & \textbf{Real Audio} & \textbf{Deepfake Audio} \\
\hline
\textbf{Total Samples} & 13,796 & 12,796 \\
\hline
\textbf{Audio Format} & WAV & WAV \\
\hline
\textbf{Sample Duration} & 6--7 seconds & 6--7 seconds \\
\hline
\textbf{Sampling Rate} & 22,050 Hz & 22,050 Hz \\
\hline
\textbf{Speaker Diversity} & 7 Speakers (male/female) & 1 Speaker (male) \\
\hline
\end{tabular}
\end{table}

\subsection{Data Composition}
The dataset is divided into two primary categories:

\textbf{Real Audio:} The real audio samples in this dataset are sourced from the SUST TTS Corpus \cite{sustTTS} and Mozilla Common Voice \cite{commonvoice:2020}. This diversity in real speech data helps to train and evaluate deep-fake detection models effectively.

\textbf{Deepfake Audio:} These are generated by a VITS-based Text-to-Speech (TTS) model trained specifically on the SUST TTS Corpus. The details of this process are discussed in the methodology section. The model generates speech that mimics real human voices, but some subtle artifacts may be present due to the limitations of the model.

The composition and audio characteristics of the dataset are summarized in Table~\ref{tab:banglafake_dataset_composition}, providing a foundation for a robust evaluation and benchmarking of Bengali deepfake detection systems.

\begin{table}[htbp]
\centering
\caption{Summary of Sources and Generated Audio Statistics}
\label{tab:banglafake_dataset_composition}
\begin{tabular}{|l|c|c|}
\hline
\textbf{Source} & \textbf{Real Audio} & \textbf{Deepfake Audio} \\
\hline
SUST TTS Corpus            & 9,999 & 9,999 \\
\hline
Mozilla Common Voice (Speaker 1) & 918  & 918 \\
\hline
Mozilla Common Voice (Speaker 2) & 573  & 573 \\
\hline
Mozilla Common Voice (Speaker 3) & 537  & 537 \\
\hline
Mozilla Common Voice (Speaker 4) & 420  & 420 \\
\hline
Mozilla Common Voice (Speaker 5) & 349  & 349 \\
\hline
Bengali News Dataset       & 1,000 & -- \\
\hline
\textbf{Total}  & \textbf{13,796} & \textbf{12,796} \\
\hline
\end{tabular}
\end{table}

\subsection{Dataset Structure}
We structured the dataset using the \textbf{LJ Speech} format \cite{ljspeech17} to maintain consistency and compatibility with existing tools. This format organizes the data with metadata files that map text to audio and standardized naming conventions, facilitating easy integration with TTS models and related frameworks. The LJ Speech format is widely adopted because of its simplicity, scalability, and ability to handle diverse datasets, making it an ideal choice for our dataset. Using this format ensures that our dataset can be readily used for training, evaluation, and comparison across different deepfake detection and generation systems.

\subsection{Dataset Usage}
The \textbf{BanglaFake} dataset is intended for various deepfake audio detection tasks, including:

\begin{itemize}
    \item \textbf{Training Deepfake Detection Models}: The dataset can be used to train machine learning models to classify real and fake audio samples.
    \item \textbf{Benchmarking Detection Techniques}: Researchers can use the dataset to benchmark the performance of different deepfake audio detection methods.
    \item \textbf{Cross-Language Comparisons}: Although focused on Bengali, the dataset could also serve as a basis for comparing deepfake detection models across different languages, especially for underrepresented languages.
\end{itemize}

\subsection{Dataset Availability}
Our dataset is publicly available at \url{https://huggingface.co/datasets/sifat1221/banglaFake}, with instructions for downloading and using it. It is released under an open license to support deepfake detection research in low-resource languages like Bengali.

\section{Methodology}
In this section, we discuss the detailed methodology of training our model with the phonetically balanced SUST TTS Bangla-speech dataset. Later we explain how we used the trained model to create the Bangla Deepfake audio dataset. Finally, we complete the discussion by evaluating the quality of the generated Deepfake audio.

\begin{figure*}[t]
  \centering
  \includegraphics[width=\textwidth]{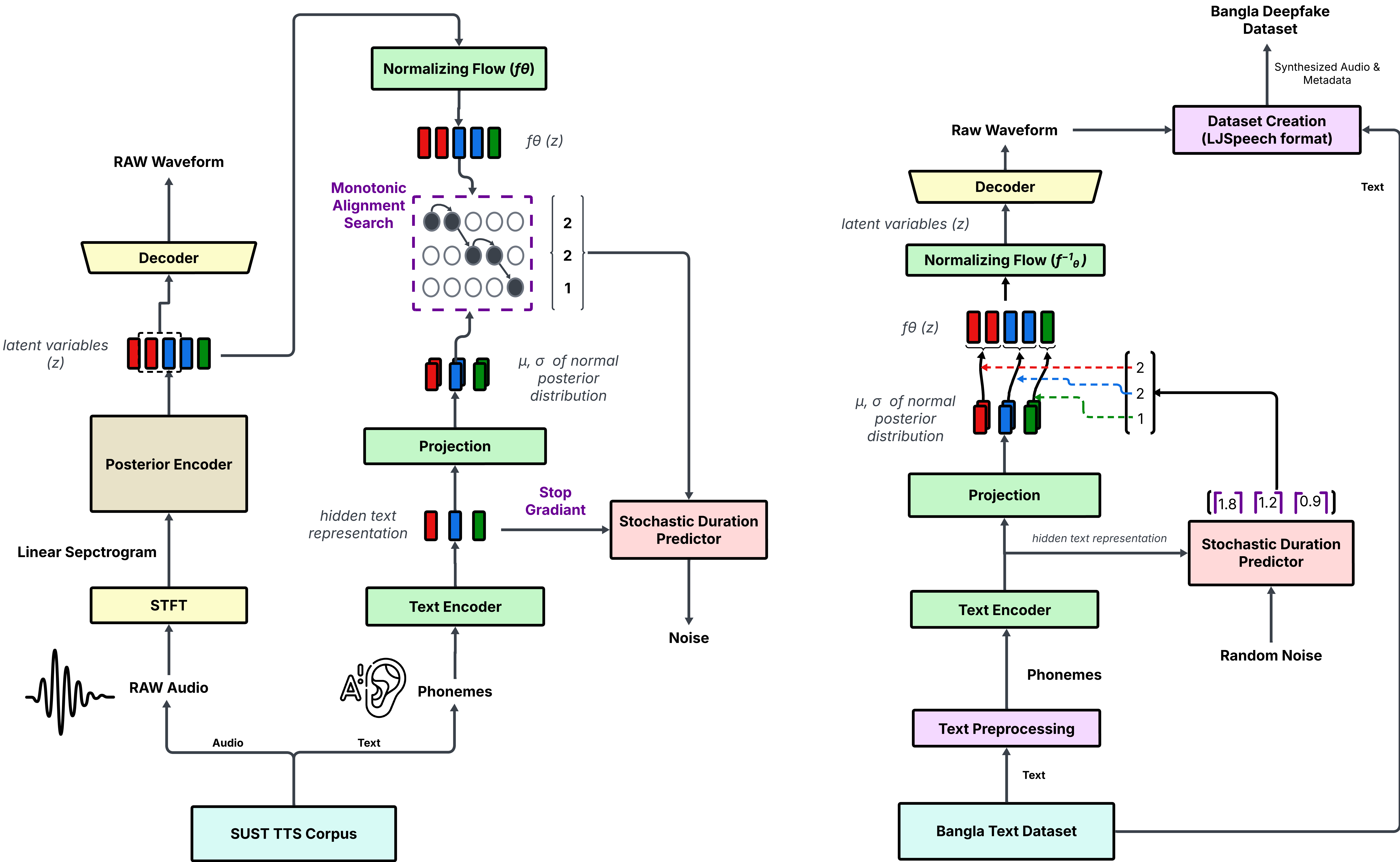}
  \caption{Overview of the Bangla Fake Speech Dataset Generation Pipeline. The left side illustrates the process of extracting latent speech representations from the SUST TTS corpus using a posterior encoder, normalizing flow, and stochastic duration predictor. The right side depicts the synthesis process, where text from the Bangla dataset is encoded and converted into speech-like representations using a trained model. The final synthesized speech dataset follows the LJSpeech format.}
  \label{fig:banglafake_methodology}
\end{figure*}

\subsection{Model Training}
We used a parallel end-to-end Text-to-Speech (TTS) method namely VITS \cite{kim2021conditional} to make a model that can generate synthetic audio from real audio. As a dataset, we have utilized the SUST TTS Corpus \cite{sustTTS}, a phonetically balanced speech dataset specifically designed for the synthesis of Bangla speech. You can find the diagram of the training procedure in Fig.~\ref{fig:banglafake_methodology}. Here, we briefly explain the VITS methodology in the following section as an overview of the process.

The training of VITS follows an end-to-end approach that combines variational inference \cite{ganguly2021introduction} and adversarial learning to generate natural-sounding speech from text. The process starts with converting text into phonemes and aligning them with latent variables using Monotonic Alignment Search (MAS) \cite{kim2020glow}. During text processing, the model also takes the raw audio as input and transforms it into a linear spectrogram using the Short-Time Fourier Transform (STFT) \cite{allen1977short}. A posterior encoder then extracts latent variables from the target waveform's linear-scale spectrogram, while a prior encoder refines their distribution. The HiFi-GAN-based \cite{kong2020hifi} decoder then up-samples these variables to generate raw waveforms, with an L1 loss computed against target mel spectrograms.

A stochastic duration predictor models phoneme durations using a flow-based generative approach, with variational dequantization \cite{ho2019flow++} improving accuracy. Adversarial training enhances speech quality through a multi-period discriminator, optimizing the generator with a feature-matching loss. The final objective combines reconstruction loss, KL divergence, duration loss, and adversarial losses. Windowed generator training improves efficiency, and mixed-precision training with AdamW optimizer ensures stable learning. This approach enables VITS to produce natural and expressive speech. Fig.~\ref{fig:banglafake_methodology} illustrates the overall working procedure including training the model and generating fake audio following the VITS method.

\subsection{Fake Audio Generation}
The fake audio generation process begins with a sequence of phonemes derived by \textbf{pre-processing} the Bangla Text Dataset. This sequence is processed by a \textbf{Text Encoder} to produce a \textit{hidden text representation} that conditions a \textbf{stochastic duration predictor}, which samples the duration for each phoneme. These durations are converted to integers and used to align phonemes and latent variables. The Text Encoder generates the mean and variance for the prior distribution, which is then transformed by a \textbf{Normalizing Flow} for increased flexibility. The alignment and phoneme sequence created a conditional prior distribution, which is used by the \textbf{Decoder} to generate a raw waveform. The decoder uses transposed convolutions and multi-receptive field fusion modules to produce the final audio output. Then we converted these audio and corresponding text representations into LJSpeech format which ultimately created the Bangla Deepfake Audio corpus. The details of this procedure can be found in Fig.~\ref{fig:banglafake_methodology}.

We utilized a total of \textbf{9,999} text samples and their corresponding real audio recordings from the SUST TTS corpus. Additionally, we include \textbf{2,797} text and real audio samples from the Mozilla Common Voice dataset \cite{ardila2019common} for five speakers to represent voice conversion scenarios, the bonafide audios being the original recordings and the fake audios generated using our model. The Mozilla subset comprises 918, 573, 537, 420, and 349 samples for Speakers 1--5, respectively. To improve speaker diversity in the real speech portion of the corpus, we further added \textbf{1,000} bonafide recordings from a Bengali news dataset, for which no synthetic counterparts were generated. Table~\ref{tab:banglafake_dataset_composition} summarizes the resulting source-wise composition: \textbf{13,796} real and \textbf{12,796} deepfake utterances in total.

\begin{figure}[htbp]
    \centering
    \includegraphics[width=0.45\textwidth]{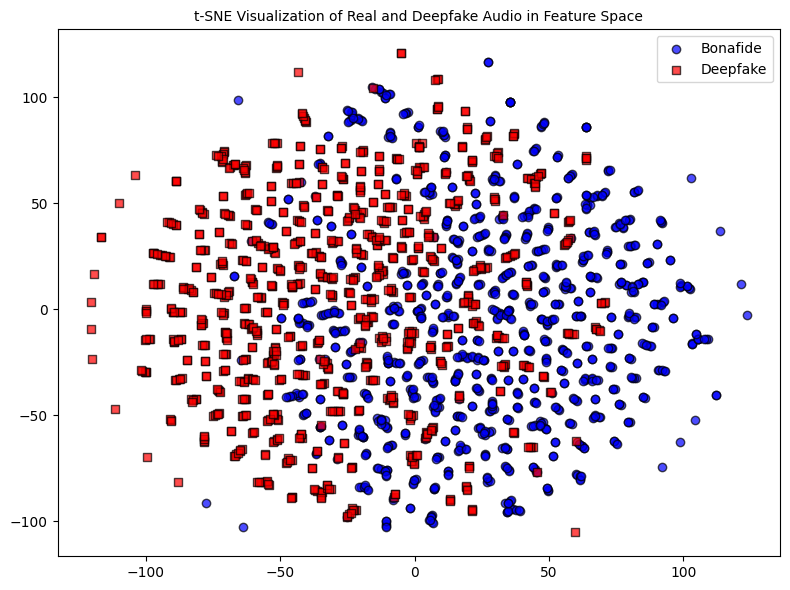}
    \caption{t-SNE visualization of MFCC features for real and deepfake audio samples. Real audio samples are represented in blue circles, while deepfake audio samples are shown in red squares. The overlap between the two classes highlights the challenge of distinguishing fake audio from real audio.}
    \label{fig:tsne_plot}
\end{figure}

\section{Evaluation}
Our dataset was evaluated through qualitative and quantitative measures to assess the effectiveness and robustness of the dataset for deepfake audio detection.

\subsection{Qualitative Evaluation}
We conducted a qualitative evaluation to assess the realism and perceptual quality of our dataset. For qualitative assessment, the following criteria were considered:

\subsubsection{Mean Opinion Score (MOS)}
Mean opinion score (MOS) \cite{mos} is one of the most popular metrics for subjective evaluation. MOS measured the naturalness of a TTS system. Five labels are given to some users for scoring the system. These five labels are Excellent, Good, Fair, Poor, and Bad. After listening to the waveforms, the individual user chooses a label to rate each of the synthesized waveforms. These five labels are mapped with some pre-defined numbers ranging from 5 to 1.

A total of \textbf{30} native Bangla speakers volunteered as listeners for the evaluation. The participants were between 20 and 25 years old, with 18 men and 12 women.

For evaluation, we synthesized speech from 10 randomly selected sentences outside of our training corpus. We designed a survey form to assess the quality of generated fake audio in a controlled experiment. Participants listened to five sets of real and fake audio clips and provided feedback based on these three questions:

\begin{enumerate}
    \item Does the fake speech sound natural and human-like?
    \item Can you clearly understand the spoken content in the fake audio?
\end{enumerate}

To ensure a robust and unbiased evaluation, we excluded the highest and lowest scores for each question when calculating the MOS, referred to as the Robust MOS. The average Robust-MOS scores for these questions were \textbf{3.40}, and \textbf{4.01}, respectively.

These results confirm that our system achieves \textbf{acceptable performance} in terms of \textbf{quality}, \textbf{naturalness}, and \textbf{intelligibility of the generated audio}. Both the subjective evaluation and objective tests demonstrate the effectiveness of our approach. Notably, our single-speaker model was trained on a carefully curated 30-hour phonetically balanced dataset. The results validate the importance of using a high-quality dataset to achieve state-of-the-art performance for Bangla TTS systems.

\subsection{Quantitative Evaluation}
The performance of deepfake detection models on the Bengali Deepfake dataset was quantitatively evaluated using several standard metrics. These metrics are crucial for objectively measuring the models' ability to distinguish between bonafide and deepfake audio samples. We evaluated the models using the following metrics:




\subsubsection{t-SNE Visualization}
To analyze the separability of real and deepfake audio, we visualize the feature space using t-SNE (t-Distributed Stochastic Neighbor Embedding) \cite{tsne}. We randomly selected \textbf{1,000 pairs} of real and deepfake audio samples from the dataset for this analysis. We extract Mel-Frequency Cepstral Coefficients (MFCCs) from both real and deepfake audio samples and reduce the high-dimensional feature space to 2D for visualization.

The t-SNE plot Fig.~\ref{fig:tsne_plot} reveals the clustering patterns of real and deepfake audio, highlighting the challenge of distinguishing between the two classes due to significant overlap in the feature space. This overlap underscores the high naturalness and quality of the synthetic speech in our dataset.

\section{Conclusion}

In this paper, we introduced the \textbf{Bengali Deepfake Audio Dataset} for a single speaker (male), a resource designed to advance deepfake audio detection for Bangla. By employing state-of-the-art Text-to-Speech (TTS) techniques, we ensured that the synthetic audio retains high levels of \textbf{naturalness} and \textbf{quality}, making detection a challenging task. Previously, we only had a Bengali TTS dataset but no dataset for fake Bengali audio. Therefore, this dataset will serve as a pioneer for developing an audio spoof detection system for Bangla. Expanding the dataset by incorporating multiple speakers (male and female voices) for better diversity and clarity and including phonemes of complex Bengali words to enhance its robustness are kept for future works.


\bibliographystyle{IEEEtran}
\bibliography{mybib}

\end{document}